\documentclass[apj]{emulateapj}
\slugcomment{{\sc Accepted for publication in ApJL:} May 20, 2015}

\usepackage{graphicx}
\usepackage{epstopdf}
\usepackage{stmaryrd}
\usepackage{amsmath}
\usepackage{ccaption}
\usepackage{multirow}
\usepackage{bm}
\usepackage{nicefrac}
\usepackage{comment}
\usepackage{textcomp}
\usepackage{color}
\definecolor{red}{rgb}{1,0,0}

\bibpunct[; ]{(}{)}{,}{a}{}{,}

\newcommand{\msun}{\ensuremath{M_{\odot}}}
\newcommand{\mstar}{\ensuremath{M_{\star}}}

\newcommand{\bdm}{\begin{displaymath}}
\newcommand{\edm}{\end{displaymath}}
\newcommand{\beq}{\begin{equation}}
\newcommand{\eeq}{\end{equation}}
\newcommand{\bit}{\begin{itemize}}
\newcommand{\eit}{\end{itemize}}
\newcommand{\ben}{\begin{enumerate}}
\newcommand{\een}{\end{enumerate}}
\newcommand{\bfi}{\begin{figure}[htb]}
\newcommand{\bpfi}{\begin{figure}[p]}

\newcommand{\lco}{\ensuremath{L'_{\rm CO}}}
\newcommand{\aCO}{\ensuremath{\alpha_{\rm CO}}}

\newcommand{\mmol}{\ensuremath{M_{\rm mol.}}}

\shorttitle{Gas fraction in a massive $z$\,=\,1.43 elliptical galaxy}
\shortauthors{Sargent et al.}

\begin{document}


\title{A direct constraint on the gas content of a massive, passively evolving elliptical galaxy at $z$\,=\,1.43$^\S$}

\author{M.~T. Sargent\altaffilmark{1, 2, $\star$},
E. Daddi\altaffilmark{2},
F. Bournaud\altaffilmark{2},
M. Onodera\altaffilmark{3},
C. Feruglio\altaffilmark{4, 5, 6},
M. Martig\altaffilmark{7, 8},
R. Gobat\altaffilmark{2, 9},
H. Dannerbauer\altaffilmark{10},
E. Schinnerer\altaffilmark{7}
}

\altaffiltext{$\star$}{~E-mail: \texttt{Mark.Sargent@sussex.ac.uk}}
\altaffiltext{1}{~Astronomy Centre, Department of Physics and Astronomy, University of Sussex, Brighton, BN1 9QH, UK}
\altaffiltext{2}{~CEA Saclay, DSM/Irfu/S\'ervice d'Astrophysique, Orme des Merisiers, F-91191 Gif-sur-Yvette Cedex, France}
\altaffiltext{3}{~Institut f\"ur Astronomie, Departement f\"ur Physik, ETH Z\"urich, Wolfgang-Pauli-Strasse 27, CH-8093 Z\"urich, Switzerland}
\altaffiltext{4}{~IRAM, 300 rue de la Piscine, F-38406 St. Martin d'Heres, Grenoble, France}
\altaffiltext{5}{~Scuola Normale Superiore, Piazza dei Cavalieri 7, I-56126 Pisa, Italy}
\altaffiltext{6}{~INAF - Osservatorio Astronomico di Roma, via Frascati 33, 00044 Monte Porzio Catone (RM) Italy}
\altaffiltext{7}{~Max-Planck-Institut f\"ur Astronomie, K\"onigstuhl 17, D-69117 Heidelberg, Germany}
\altaffiltext{8}{~Centre for Astrophysics \& Supercomputing, Swinbourne University, Melbourne, Australia}
\altaffiltext{9}{~KIAS, 85 Hoegiro, Dongdaemun-gu Seoul 130-722, Republic of Korea}
\altaffiltext{10}{~Institut f\"ur Astronomie, Universit\"at Wien, T\"urkenschanzstrasse 17, A-1160 Wien, Austria}
\altaffiltext{$\S$}{~Based on observations with the Plateau de Bure millimetre Interferometer (PdBI), operated by the Institute for Radio Astronomy in the Millimetre range (IRAM), which is funded by a partnership of INSU/CNRS (France), MPG (Germany) and IGN (Spain).}

\begin{abstract}
Gas and dust in star-forming galaxies at the peak epoch of galaxy assembly are presently the topic of intense study, but little is known about the interstellar medium (ISM) of distant, passively evolving galaxies. We report on a deep 3\,mm-band search with IRAM/PdBI for molecular (H$_2$) gas in a massive ($M_{\star}$\,$\sim$\,6$\times$$10^{11}$\,$M_{\odot}$) elliptical galaxy at $z$\,=\,1.4277, the first observation of this kind ever attempted. We place a 3\,$\sigma$ upper limit of 0.32\,Jy\,km/s on the flux of the CO($J$=2$\rightarrow$1) line or $L'_{\rm CO}$\,$<$\,8.8$\times$10$^{9}$\,K\,km/s\,pc$^2$, assuming a disk-like CO-morphology and a circular velocity scaling with the stellar velocity dispersion as in local early-type galaxies (ETGs). This translates to an H$_2$ mass of $<$3.9$\times$10$^{10}$\,($\alpha_{\rm CO}$/4.4)\,$M_{\odot}$ or a gas fraction of $\lesssim$6\,\% assuming a Salpeter initial mass function (IMF) and an ISM dominated by H$_2$, as observed in many local, high-mass ellipticals. This low value approaches that of local ETGs, suggesting that the low star formation activity in massive, high-$z$ passive galaxies reflects a true dearth of gas and a lesser role for inhibitive mechanisms like morphological quenching.\end{abstract}


\section{Introduction}
\label{sect:intro}

Molecular gas has been detected through rotational CO transitions out to very high redshift, first in highly luminous sources experiencing ``bursty" star formation \citep[sub-millimetre galaxies \& QSOs; e.g.][]{greve05, riechers06, maiolino07} and subsequently in galaxies undergoing ``main-sequence" (MS) star formation \citep[SF; e.g.][]{daddi08, tacconi10, geach11}. Observations show that, when the universe was half its current age, MS galaxies had gas reservoirs approximately 10-fold larger than nearby spiral galaxies \citep[e.g.][]{daddi10a, tacconi13}.\\
It is not known, however, whether the gas content evolves in a synchronised fashion in both active and passive galaxies, i.e. following a ``universal" cosmic decline. Molecular gas is detected in roughly \nicefrac{1}{5} of all local ETGs \citep[slightly more in the field than in clusters; cf.][]{young11} and typically represents $\sim$1\% of the stellar mass of morphological types E/S0. Do ETG gas fractions increase with redshift? Intuition might suggest this as massive high-$z$ ETGs undergo mergers with gas-rich satellites and accrete a fraction of the latter's gas. \citet{gabor10} show that standard quenching mechanisms (gas consumption in mergers, AGN feedback, virial shocks) produce red sequence galaxies with bluer colors than observed at $z$\,$=$\,1-2 because their simulated spheroids retain substantial residual gas, resulting in excessive SF unless additional quenching mechanisms are invoked. Fragmentation-prone gas reservoirs may be kept from forming stars when embedded within a stellar spheroid \citep[``morphological quenching";][]{martig09}, implying that high-$z$ ETGs could host gas which cannot be tapped to form stars and has so far escaped attention. This is also consistent with the long gas consumption time scales \citet{saintonge12} find in nearby ETGs.\\
To obtain a first direct measurement of the currently unknown molecular gas reservoirs of distant passive galaxies we searched for CO($J$=2$\rightarrow$1) line emission from a massive elliptical galaxy at $z$\,$\sim$\,1.5 with the PdBI. Our target and observations are described in Section \ref{sect:obs}. Results and discussion follow in Sections \ref{sect:results} and \ref{sect:discussion}. We adopt a flat WMAP-7 cosmology \citep[$\Omega_m$\,=\,0.273 and $H_0$\,=\,70.4 km\,s$^{-1}$\,Mpc$^{-1}$;][]{larson11}.\newpage

\section{Target description and data}

\subsection{J100239.52+015659.1 - a pBzK-selected, massive, elliptical galaxy at $z$\,=\,1.4277}
\label{sect:target}

A stringent limit on the gas content of distant quiescent galaxies allowing a meaningful comparison with nearby ETGs is most readily achieved for high-mass galaxies. We selected the most massive galaxy -- J100239.52+015659.1 (pBzK-217431) -- of \citet[``O12" hereafter]{onodera12}. pBzK-217431 is a BzK-selected \citep{daddi04}, passively evolving, red elliptical galaxy in the COSMOS field with spectroscopic redshift\footnote{~The redshift error translates to a $\pm$0.6\,GHz frequency uncertainty ($\pm$5\,$\sigma$) of the redshifted CO($J$=2$\rightarrow$1) line, well within the 3.6\,GHz instantaneous bandwidth of the WideX correlator used here.}
$z$\,=\,1.4277$\pm$0.0015 (derived from a high-confidence detection of the 4000\,{\AA} break) and age $\sim$1.1\,Gyr \citep[corresponding to the average `build' redshift $z$\,$\sim$\,2.3 for the bulk of stars in ETGs in the O12 sample;][]{onodera15}.
With a stellar mass\footnote{~In accordance with literature on IMF variations \citep[e.g.][]{grillogobat10, conroydokkum12, cappellari12} we henceforth adopt a \citet{salpeter55} IMF for pBzK-217431 and a \citet{chabrier03} IMF for SFGs.} of 6.6$^{+0.5}_{-1.9}$\,$\times$\,$10^{11}$\,$M_{\odot}$ (from SED-fitting in O12) this object is among the most massive and luminous ($K_{\rm Vega}$\,=\,17.56) ETGs known at this redshift. It is 5-10 times more massive than star-forming galaxies (SFGs) at a similar redshift followed-up in CO by \citet{daddi10a} or the PHIBSS survey \citep{tacconi13}. pBzK-217431 lies within 1\,$\sigma$ of the local mass-size relation for elliptical galaxies \citep[e.g.][]{newman12} and has an optical surface brightness distribution consistent with the canonical de Vaucouleurs profile \citep[O12,][]{mancini10}. Stellar population modelling and SED-fitting indicate that its star formation rate (SFR) is $\lesssim$6\,$M_{\odot}$/yr \citep{onodera15}.

\subsection{IRAM PdBI 3\,mm-band CO($J$=2$\rightarrow$1) follow-up}
\label{sect:obs}

\begin{figure}
\epsscale{.94}
\centering
\plotone{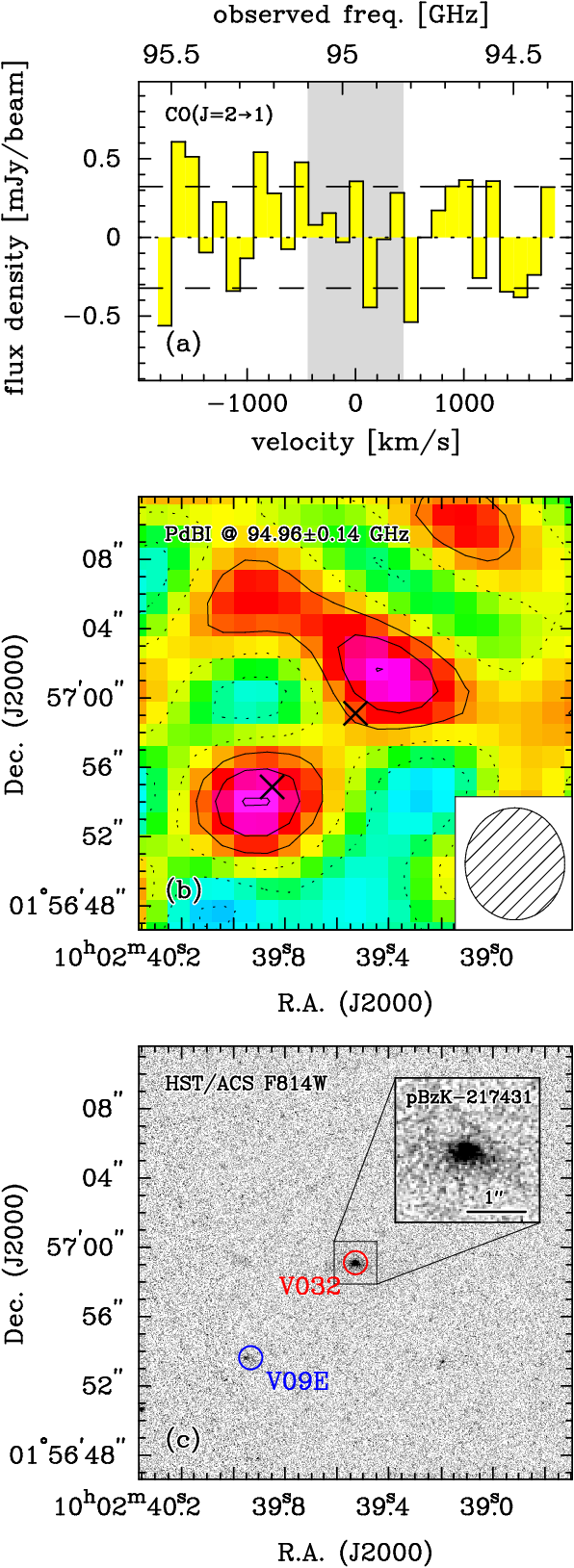}
\caption{PdBI 3\,mm observations of pBzK-217431. {\it (a)} Spectrum extracted at the target position within $\pm$2000\,km/s of the expected frequency of redshifted CO($J$=2$\rightarrow$1). Dashed lines: $\pm$1\,$\sigma$ noise level for 40\,MHz channels. {\it (b)} Map of the emission integrated over the expected FWHM of the CO line (280\,MHz or 882\,km/s; see panel {\it (a)}, grey area). Solid (dotted) contours: positive (negative) 1 \& 2\,$\sigma$ noise level. Inset: synthesized beam. {\it (c)} HST $I$-band imaging of pBzK-217431 and surroundings (image size -- 30$''$$\times$30$''$; zoom-in size -- 2.5$''$$\times$2.5$''$). Blue/red circle: phase centre of the two separate PdBI pointings targeting the pBzK-217431 field (Section \ref{sect:obs}). Pointing centres are indicated with crosses in panel {\it (b)}.
\label{fig:basics}}
\end{figure}
pBzK-217431 was observed in several tracks between June 21-30, 2011 (IRAM project ID {\it V032}) and May 26-July 19, 2012  (project ID {\it V09E}), always with five PdBI antennae on-source, excluding one track in 2012 conducted with only four antennae. During both observation campaigns all tracks were carried out in the compact D-configuration (angular resolution: 6.5$''{\times}$5.7$''$), with the 3.6\,GHz WideX correlator tuned to the expected frequency of redshifted CO($J$=2$\rightarrow$1) at 94.96\,GHz ($\nu_{\rm rest}$\,=\,230.538\,GHz). In 2011 the phase centre of the observations lay on pBzK-217431 and in 2012 on a nearby dwarf-like galaxy with photometric redshift $z_{\rm phot}$\,=\,0.215 \citep{ilbert13}, offset $\sim$7$''$ from pBzK-217431 (Figure \ref{fig:basics}c). This shift was applied following the detection of a broad (760\,MHz; detected with $S/N$\,=\,6.5 in the $uv$-plane) emission feature associated with the neighboring galaxy in multiple tracks in 2011. Given the photometric redshift of the dwarf galaxy this could have represented a CO($J$=1$\rightarrow$0) line emission feature but the data gathered in 2012 suggested this was an artefact. We reduced the data from the two observing periods individually with the GILDAS software packages CLIC and MAPPING and relying on varying combinations of the calibrators 0851+202, 0906+015, 0923+392, 1005+066, 1040+244, 1055+018, 3C84, 3C273, 3C345, CRL618 and MWC349 for pointing and phase calibration, and on either 0923+392, 1005+066, 3C279, MWC349, 3C84 or 3C345 for flux calibration. Data sets {\it V032} and {\it V09E} were subsequently combined after correcting for primary beam attenuation and phase shifting between the two pointing centers. After flagging of poor visibilities the combined data set has a total 5-antenna (6-antenna) equivalent observing time of 11.59 (7.73)\,hr and reaches a noise level of 0.33\,mJy/beam per 40\,MHz channel (126\,km/s). This sensitivity was not sufficient to detect CO emission. Figure \ref{fig:basics}a shows the featureless spectrum binned in 40\,MHz channels, extracted at the location of pBzK-217431 and centred on the expected position of redshifted CO($J$=2$\rightarrow$1).

\section{Results}
\label{sect:results}

To set an upper limit on the CO($J$=2$\rightarrow$1) line flux we assume that -- as observed in most massive, local ETGs \citep{alatalo13} -- the molecular gas in pBzK-217431 has settled into a rotating disk with circular velocity $V_c\,=\,\kappa\sigma_{\star}$. Following the dynamical analysis on ATLAS$^{\rm 3D}$ ETGs by \citet{cappellari13}, we set $\kappa\,{\sim}\,1.64$ (i.e. intermediate between the maximal circular velocity and the circular velocity at the effective radius). We have no direct measurement of the velocity dispersion for pBzK-217431 but adopt the average\footnote{~A similar line width is inferred by assuming that velocity dispersion, mass and size are related as $\sigma^2_{\star}$\,$\propto$\,\mstar/$r_e$ and estimating the velocity dispersion of pBzK-217431 (\mstar\,=\,6.6$\times$10$^{11}$\,$M_{\odot}$, $r_e$\,=\,7.2\,kpc) based on the O12 sample median [$\sigma_{\star}$, \mstar, $r_e$]\,=\,[330\,km/s, 2.2$\times$10$^{11}$\,$M_{\odot}$, 2.4\,kpc]. This re-scaling procedure involving size measurements is only meaningful when all galaxies have similar structural properties; the median S\'ersic index and scatter of the O12 sample is 3.2$\pm$1.2, implying that pBzK-217431 lies within 0.5\,$\sigma$ of the sample median.} $\sigma_{\star}$\,=\,330\,km/s of the 1.4\,$<$\,$z$\,$<$\,2.1 ETG parent sample \citep{onodera15}. This value agrees well with velocity dispersions found by other high-$z$ ETG studies \citep[e.g.][]{newman10} and translates to an expected line width $2V_c\,{\sim}$\,910\,km/s (approx. seven spectral channels) if we assume a gas disk with inclination angle\footnote{~Line width and velocity-integrated flux $I_{\rm CO}$ vary with viewing angle as ${\rm sin}(i)$ and $\sqrt{{\rm sin}(i)}$, respectively. For an edge-on gas disk the upper limits on the CO-luminosity and H$_2$ mass would thus increase by $\sim$10\%.} $i\,{=}\,57.3^{\circ}$, i.e. the mean orientation for randomly oriented disks in 3D.\\
Figure \ref{fig:basics}b shows the flux distribution within $\sim$15$''$ of the target position after integration of the 3\,mm spectrum over seven channels (280\,MHz; 882\,km/s in velocity space) centred on the expected line frequency. While we see a $\sim$3\,$\sigma$ flux concentration 3$''$ north of pBzK-217431 in Figure \ref{fig:basics}b we do not consider this a reliable signature of weak line emission from our target. This is supported by the fact that the total number of $>$2\,$\sigma$ pixels in the displayed region is fully consistent with Gaussian statistics. At very low signal-to-noise the most robust estimate of the flux is generally obtained at the phase centre where we detect flux only at 1.2\,$\sigma$ significance. The 1\,$\sigma$ noise level for $I_{{\rm CO}(J{=}2{\rightarrow}1)}$ is 0.11\,Jy\,km/s, leading to a 3\,$\sigma$ upper line luminosity limit $L'_{{\rm CO}(J{=}2\rightarrow1)}$\,$<$\,8.8$\times$10$^{9}$\,K\,km/s\,pc$^2$ following \citet[equation 3]{solomonvandenbout05}.\\
For ATLAS$^{\rm 3D}$ ETGs CO($J$=2$\rightarrow$1) emission is consistent with thermal excitation \citep{young11} and can thus be used as a direct tracer of the H$_2$ content. Converting our line flux limit to a constraint on the gas mass of pBzK-217431 requires choosing a CO-to-H$_2$ conversion factor. ISM conditions in low-$z$ ETGs generally resemble those of local spiral galaxies (dense gas fractions -- \citealp{krips10}; H$_2$ surface densities -- \citealp{young11}; dust temperatures -- \citealp{crocker12}; low star formation efficiency, SFE -- \citealp{martig13}), leading us to adopt a Milky-Way-like \aCO\,$\sim$\,4.4\,$M_{\odot}$/(K\,km/s\,pc$^2$). A similar \aCO\ is inferred when considering a metallicity-dependent conversion factor \citep[e.g.][]{schruba12, genzel12} and using the slightly super-solar, average enrichment $Z/Z_{\odot}$\,$\sim$\,1.2 of ETGs of the O12 sample. Finally, we note that simulations of high-$z$ ellipticals with a range of gas fractions ($\sim$10-50\%) and post-processing with Large Velocity Gradient (LVG) models predict \aCO-values comparable to that of the Milky Way as well \citep{bournaud14}.\\
\begin{deluxetable}{lr}
\tabletypesize{\small}
\tablewidth{0pt}
\tablecaption{Source properties: pBzK-217431. \label{tab:prop}}
\tablehead{
\colhead{quantity/observable} &
\colhead{value}}
\startdata
R.A. [J2000] & 10$^h$02$^m$39.527$^s$\\
Dec. [J2000] & +01$^d$56$^m$59.12$^s$\\
$z_{\rm spec}$ & 1.4277$\pm$0.0015\\
\mstar\ [$M_{\odot}$] & 6.6$^{+0.5}_{-1.9}$\,$\times$\,$10^{11}$\\
$r_e$ [kpc] & 7.19$\pm$1.95\\
S\'ersic index $n$ & 3.8$\pm$0.6\\[0.5ex]
\hline\\[-2ex]
rms/40\,MHz [mJy] & 0.33\\
$I_{\rm CO(J{=}2\rightarrow1)}$ [Jy\,km/s] & $<$0.32\,$\sqrt{\left(\frac{\Delta v}{882\,{\rm km/s}}\right)}$\\
$L'_{\rm CO(J{=}2\rightarrow1)}$ [K\,km/s\,pc$^2$] & $<$8.8$\times$10$^{9}$\\
\mmol\ [$M_{\odot}$] & $<$3.9$\times$10$^{10}$\,$\left(\frac{\alpha_{\rm CO}}{4.4}\right)$\\
$f_{\rm gas}$ & $<$5.8\%
\enddata
\tablecomments{Properties listed above horizontal divider are from \citet{onodera15}. Upper limits are quoted at 3\,$\sigma$ significance.}
\end{deluxetable}
\begin{figure*}
\centering
\begin{tabular}{ccc}
\includegraphics[width=0.46\textwidth]{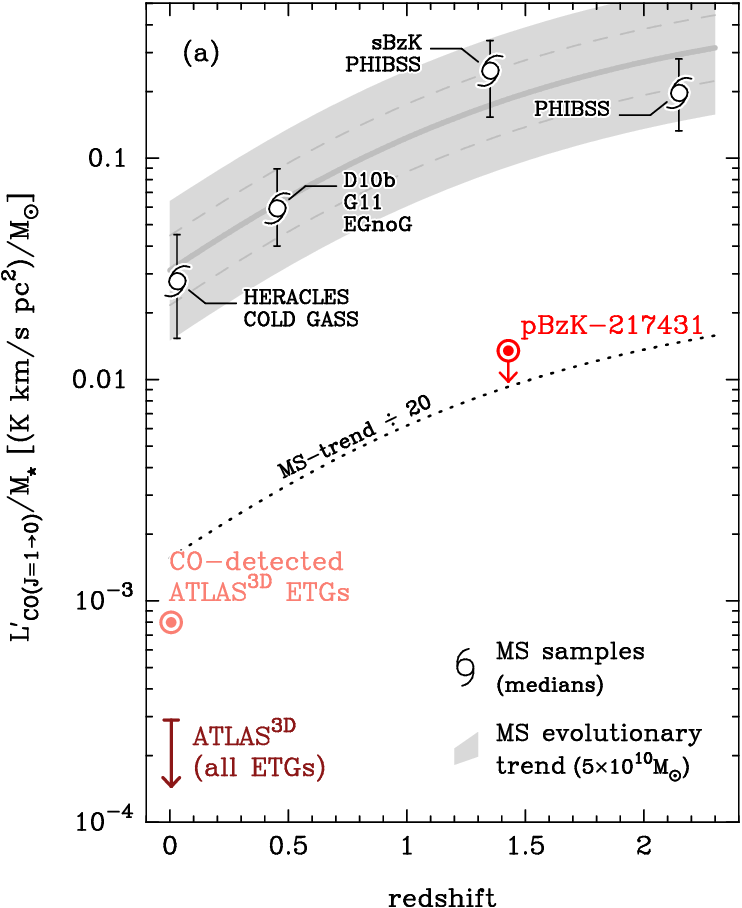} & & \includegraphics[width=0.46\textwidth]{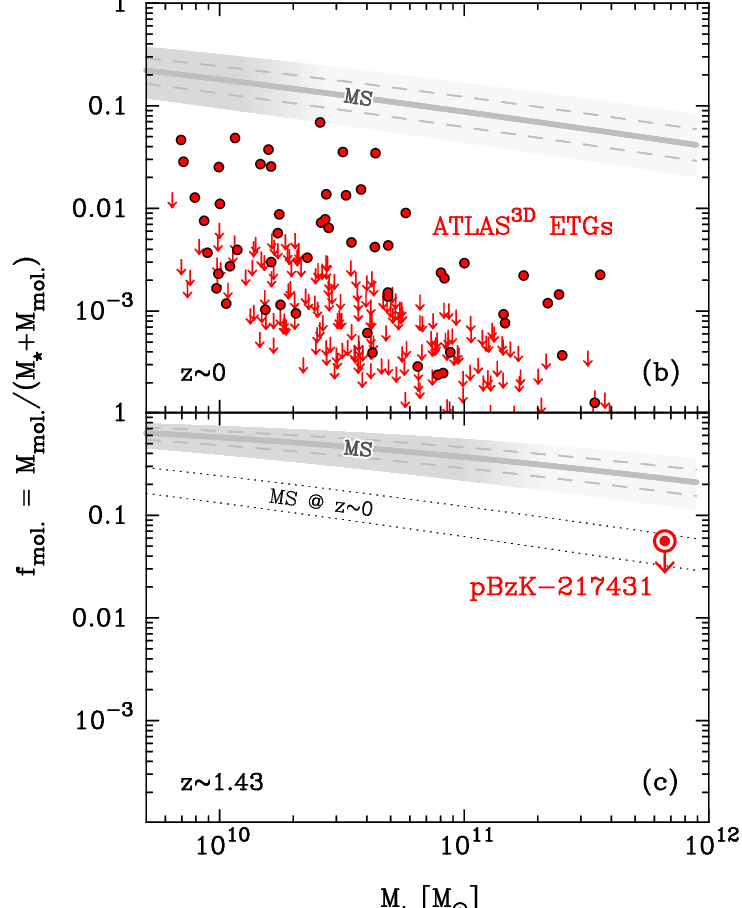}\\
\end{tabular}
\caption{CO emission and gas content of pBzK-217431 compared to ATLAS$^{\rm 3D}$ ETGs and 0\,$<$\,$z$\,$<$\,2.5 main-sequence (MS) galaxies. {\it (a)} 3\,$\sigma$ upper limit on the ratio of CO luminosity and stellar mass {\mstar} for pBzK-217431, plus typical $L'_{\rm CO}/\mstar$ ratios measured for ATLAS$^{\rm 3D}$ ETGs (light red -- median of CO-detected ETGs; dark red -- upper limit on sample median for {\it all} ATLAS$^{\rm 3D}$ ETGs) and literature MS galaxy samples. (See annotations for survey names. Author abbreviations: D10b -- \citealt{daddi10b}; G11 -- \citealt{geach11}). For MS galaxies we plot the median $L'_{\rm CO}/\mstar$ and redshift of the combined samples; error bars illustrate the 1\,$\sigma$ scatter in the data. Dark grey, solid line: expected redshift evolution of $L'_{\rm CO}/\mstar$ for an average MS galaxy with \mstar\,=\,5$\times$10$^{10}$\,\msun\ (Section \ref{sect:discussion_evo}). Dashed lines (edges of light grey shading): $L'_{\rm CO}/\mstar$ ratio for $\pm$1\,$\sigma$ ($\pm$2\,$\sigma$) outliers to the MS.\newline 
{\it (b)} Molecular gas fraction $f_{\rm mol.}$ for ATLAS$^{\rm 3D}$ ETGs compared to the average relation between gas fraction and \mstar\ for massive $z$\,$\sim$\,0 MS galaxies. (Shading/lines as in panel {\it (a)}; in panels {\it (b)} and {\it (c)} the grey shading fades away at the cross-over mass, above which ETGs dominate SFGs by number at the respective redshift.)\newline
{\it (c)} As in panel {\it (b)} but highlighting the paucity of gas in pBzK-217431 with respect to typical gas fractions of massive MS galaxies at the same redshift. Dotted lines: $\pm$1\,$\sigma$ scatter of $f_{\rm mol.}$ in $z$\,$\sim$\,0 MS galaxies (cf. panel {\it (b)}).
\label{fig:fgas}}
\end{figure*}
Atomic (HI) to molecular gas ratios in nearby ETGs vary with environment and kinematic state. Based on the fifth nearest neighbour distribution of galaxies in the COSMOS field with a similar redshift $z$\,=\,1.43$\pm$0.2 as our target, pBzK-217431 lies in an average density environment (54$^{\rm th}$ percentile of the density distribution), where HI can dominate the gas phase for nearby ETGs \citep[e.g.][]{welch10}. Conversely, it is in a mass regime where locally HI-poor slow-rotators dominate the ETG population. Absent direct HI measurements, we assume that mass is the key factor and that CO emission traces the {\it entire} gas reservoir, similar to ETGs in the local sample of \citet{crocker11}. In this case, our observations constrain the gas mass of pBzK-217431 to be $M_{\rm gas}$\,$<$\,3.9$\times$10$^{10}$\,($\alpha_{\rm CO}$/4.4)\,$M_{\odot}$. This is comparable to gas masses in MS galaxies at 1\,$<$\,$z$\,$<$\,2, whereas pBzK-217431 is an order of magnitude more massive, resulting in a 3\,$\sigma$ upper limit\footnote{~Although $M_{\rm gas}$ appears in numerator and denominator of $f_{\rm gas}$, we can nevertheless define a meaningful upper limit on the gas fraction because the inverse $\nicefrac{1}{f_{\rm gas}}$\,=1+$\nicefrac{\mstar}{M_{\rm gas}}$ represents a well defined {\it lower} limit on $\nicefrac{1}{f_{\rm gas}}$ when, as here, only an upper limit on $M_{\rm gas}$ is available.} of $f_{\rm gas}$\,=\,$\nicefrac{M_{\rm gas}}{(M_{\star}+M_{\rm gas})}$\,$<$\,5.8\%, i.e. a factor 10 lower than most CO-detected MS galaxies at the peak epoch of galaxy assembly. The properties of pBzK-217431 are summarized in Table \ref{tab:prop}.

\section{Discussion}
\label{sect:discussion}

\subsection{The gas content of pBzK-217431 in the galaxy evolution context}
\label{sect:discussion_evo}

\begin{figure*}
\epsscale{.5}
\centering
\begin{tabular}{ccc}
\includegraphics[width=0.46\textwidth]{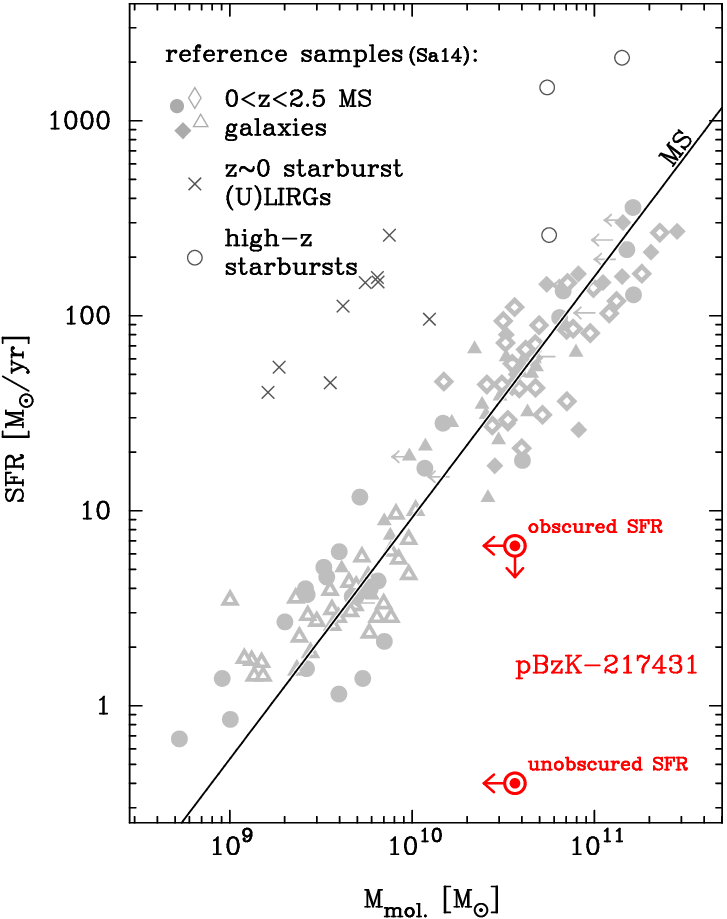} & & \includegraphics[width=0.46\textwidth]{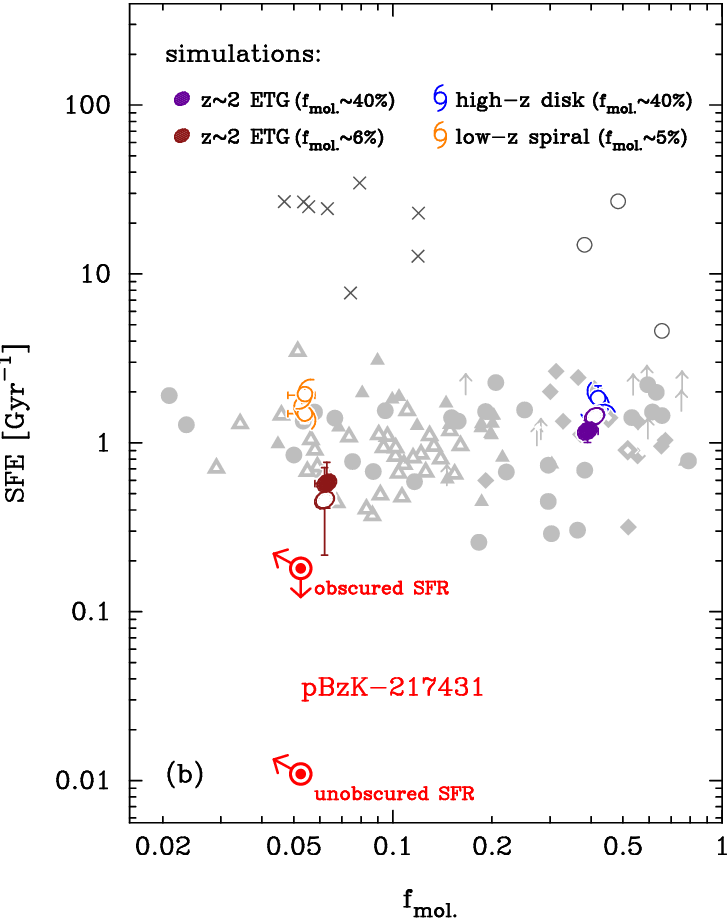}\\
\end{tabular}
\caption{Location of pBzK-217431 in the integrated Schmidt-Kennicutt diagramme {\it (a)} and in SFE vs. $f_{\rm mol.}$ space {\it (b)}. Data points scattering around the `MS' SF law in {\it (a)} are literature main-sequence galaxies \citep[compiled in][Sa14]{sargent14}. Crosses (open circles): low- (high-)$z$ starbursts with measured \aCO\ and typically $\sim$15$\times$ higher SFE than MS galaxies. Indicative SFR estimates for pBzK-217431 are from average spectral/SED properties of $z$\,$\sim$\,1.5 ETG samples \citep[O12,][]{man14}. Colored symbols indicate the properties of simulated gas-rich/gas-poor ({\it purple}/{\it dark red}) ETGs and gas-rich/gas-poor ({\it blue}/{\it orange}) disks (see discussion in text). Error bars on simulated data points illustrate variations in the simulations of these quantities over a few 100\,Myr.
\label{fig:SFE}}
\end{figure*}
As a proxy for the redshift evolution of ETG gas fractions which involves no assumptions about \aCO\ we compare in Figure \ref{fig:fgas}a the \lco/\mstar\ ratio of pBzK-217431 with sample averages for MS galaxies at $z$\,$<$\,2.5 and $z$\,$\sim$\,0 ETGs\footnote{~Masses for ATLAS$^{\rm 3D}$ ETGs are the product of the $r$-band luminosity and mass-to-light ratios from JAM modelling in \citet{cappellari12}.} from ATLAS$^{\rm 3D}$ \citep{young11}. Median \lco/\mstar\ values of literature MS galaxy samples are plotted with spiral galaxy symbols: $z$\,$\sim$\,0 -- HERACLES \& COLD GASS \citep{leroy09, saintonge11}; 0.1\,$<$\,$z$\,$<$\,0.6 -- \citet{daddi10b, geach11} \& EGNoG \citep{bauermeister13}; 1.1\,$<$\,$z$\,$<$\,1.7 \& $z$\,$>$\,2 -- \citet{daddi10a} \& PHIBSS \citep{tacconi13}. The continuous evolution of the gas content in $z$\,$<$\,2.5 MS galaxies with \mstar\,$\sim$\,5$\times$10$^{10}$\,$M_{\odot}$, the typical mass of all these samples, is traced by the grey band, based on the scaling relations between \mstar, SFR, molecular gas mass $M_{\rm mol.}$ and \aCO\ calibrated in \citet{sargent14}. For ATLAS$^{\rm 3D}$ ETGs we plot the median of the subset of CO-detected ETGs (22\% of ATLAS$^{\rm 3D}$ ETGs) and an upper limit on the median \lco/\mstar\ ratio of the overall sample, placed at the 84$^{\rm th}$ percentile of the \citet{kaplanmeier58} distribution which accounts for both the CO-detections and non-detections. The upper limit on the CO-deficit of pBzK-217431 relative to $z$\,$\sim$\,1.4 MS galaxies is $\sim$20 (see Figure \ref{fig:fgas}a), comparable to that of CO-detected ATLAS$^{\rm 3D}$ ETGs relative to local spirals. This limit is not stringent enough for a helpful comparison with the full ATLAS$^{\rm 3D}$ ETG sample median.\\ 
In the introduction we asked whether gas fractions in star-forming and quiescent galaxies evolved with time by equal proportions. A quantitative comparison must involve the fact that pBzK-217431 is about 10$\times$ as massive as ATLAS$^{\rm 3D}$ ETGs on average. Figures \ref{fig:fgas}b/c show that gas fractions of local ETGs vary inversely with \mstar, reaching $\lesssim$1\textperthousand\ at the highest masses. Our observations are not sufficiently deep to start testing whether, at fixed \mstar, $f_{\rm gas}$ for quiescent galaxies is constant over the last $\sim$10\,Gyr. Over this period gas fractions of normal galaxies with masses at or above the knee of the stellar mass function ($M^*$\,$\sim$\,10$^{11}$\,$M_{\odot}$) evolve by a factor 4-5 \citep[from a few to about 15-30\%; cf. Figure \ref{fig:fgas}b/c or][]{tacconi13, scoville14}. For pBzK-217431 and local ETGs the values are $<$6\% and $<$1\%, resp., such that we cannot exclude a similar $\sim$5-fold evolution of $f_{\rm gas}$ for ETGs based solely on our observations.\\
Combining our upper limit on the gas mass of pBzK-217431 with SFR-estimates from ETG stacked optical spectra and optical-to-IR SEDs \citep{man14, onodera15}, we can estimate its SFE. The average specific SFR of $<$\,10$^{-11}$\,yr$^{-1}$ for the ETG sample of O12 is consistent with the upper limit for the {\it dust-obscured} SFR in the largest mass bin of a larger sample of photometrically selected quiescent galaxies in \citet{man14}. Applied to pBzK-217431 this level implies a SFR of $<$5-10\,$M_{\odot}$/yr , while SED fitting \citep{man14} suggests SFR\,$\sim$\,0.4\,$M_{\odot}$/yr. The resulting, loose constraint on the gas depletion time $\tau_{\rm depl.}$\,=\,$\nicefrac{M_{\rm gas}}{\rm SFR}$ is $\lesssim$90\,($\alpha_{\rm CO}$/4.4)\,Gyr (see Figure \ref{fig:SFE}). Presuming that, similar to nearby ETGs \citep[e.g.][]{saintonge12, martig13}, high-$z$ ETGs have 2-5$\times$ lower SFEs than disk galaxies this means that a roughly 10-fold sensitivity increase compared to our PdBI observations would be required to detect an expected gas reservoir of order 10$^9$\,$M_{\odot}$. This highlights the need for alternative ISM tracers or deeper observations, e.g. with ALMA, in order to advance our knowledge of the cold gas in distant quiescent galaxies.

\subsection{Considerations on feedback and quenching}
\label{sect:discussion_feedback}

Galaxies as massive as pBzK-217431 at $z$\,=\,1.4277 are prime candidates for suffering so-called `mass quenching' \citep[e.g.][Figure 15]{peng10}. Various mechanisms potentially contribute to mass quenching, and more widespread observations of the ISM of quenched galaxies as presented here should ultimately reveal which (combination) of these is most important. Examples of internal processes are gravitational/morphological quenching \citep{martig09, genzel14}, AGN-feedback \citep[e.g.][]{granato04} or SF-activity \citep[e.g.][]{ceverinoklypin09}. Feedback suppresses SF by removing  gas, while simulations of morphological quenching have shown that the SFR of ETGs can remain minimal for several Gyr, notwithstanding gas fractions of 5-10\% \citep[cf. Figure 3 in][]{martig09}.
Our non-detection for pBzK-217431 thus indicates that massive high-$z$ ellipticals at high redshift are truly gas-poor. We illustrate this with Figure \ref{fig:SFE}b which compares SFE(s) and molecular gas fraction(s) for pBzK-217431 and low- and high-$z$ MS galaxies with simulated $z$\,$\sim$\,2 ETGs (purple -- simulated ETG with $f_{\rm mol.}$\,$\sim$\,50\%; dark red -- simulated $f_{\rm mol.}$\,$\lesssim$10\%) having \mstar\,$\sim$4$\times$10$^{10}$ (open symbols) and $\sim$10$^{11}$\,$M_{\odot}$ (filled symbols). These simulations include those from \citet{bournaud14} and similar models with other parameters (galaxy type/gas fraction) with 3\,pc resolution. They are idealized closed-box models with a chosen initial stellar structure and gas fraction, evolved for a few dynamical times including self-gravity, hydrodynamics, cooling and stellar feedback to compute the phase-space distribution of the ISM and the associated SFE. In gas-rich simulated ETGs, morphological quenching is ineffective, as evidenced by the fact that these systems have the same SFEs as both observed and simulated gas-rich disk galaxies. Conversely, morphological quenching could be responsible for the small (comparable to the dispersion of the SF-law) SFE-offset between the gas-poor simulated ETGs and simulated $z$\,$\sim$\,0 disks (orange spirals). The low SFRs of massive, high-$z$ ellipticals are thus not the result of unusually low SFE caused by gravitational stabilization of gas reservoirs, but morphological quenching might shut down SF completely in already gas-poor ETGs. Note that the absence of gas need not be the outcome of active expulsion by AGN or stellar feedback. Sub-millimeter galaxies as the commonly assumed progenitors of early ellipticals usually have sufficiently short gas consumption times \citep[e.g.][]{bothwell13} for reservoir-depletion to well below the sensitivity of our observations of pBzK-217431.

\section{Summary}
\label{sect:summary}

We presented the outcome of a deep IRAM/PdBI 3\,mm search for CO($J$=2$\rightarrow$1) emission from pBzK-217431, a massive, passively evolving elliptical galaxy at $z$\,=\,1.4277. Though undetected in CO, we could infer a 3\,$\sigma$ upper limit of 5.8\% for its gas fraction. This is almost 10$\times$ below the typical $f_{\rm gas}$ of MS galaxies at the same epoch and comes close to the gas fractions of local ATLAS$^{\rm 3D}$ ETGs, although our limit is not stringent enough for a direct comparison with the most massive ellipticals in the nearby universe where $f_{\rm gas}$\,$\sim$\,1\textperthousand. Observational data on the gas content of quiescent galaxies not only tells us about the currently unknown contribution of such galaxies to the cold gas history of the Universe \citep[e.g.][Sargent et al., in prep.]{walter14}, it also provides insight into how/which quenching and feedback processes influence the ISM of the host galaxy \citep[e.g.][]{ciottiostriker97}. Our gas fraction measurement for pBzK-217431 leads us to speculate that moderate-sized, gravitationally stabilized gas reservoirs are not widespread among massive high-$z$ ellipticals and that their low SFR reflects a paucity of gas, caused by either feedback or exhaustion of their gas reservoir by prior SF activity.

\acknowledgments
We thank J. M. Winters, A. Cibinel, P.-A. Duc, M. Krips and the referee for helpful suggestions and/or data reduction support. MTS, ED, RG \& HD acknowledge funding from ERC-StG \#240039 and grant ANR-08-JCJC-0008, FB funding from ERC-StG \#257720.

\end{document}